\documentclass{jpsj2}
\title{%
Nonlinear Optical Response Functions of Mott Insulators 
Based on Dynamical Mean Field Approximation}

\author{%
Takanobu {\sc Jujo}
\thanks{E-mail address: jujo@ms.aist-nara.ac.jp}
}

\inst{%
Graduate School of Materials Science, Nara Institute of Science 
and Technology, Ikoma, Nara 630-0101
}

\recdate{\today}

\abst{%
We investigate the nonlinear optical susceptibilities 
of Mott insulators with the dynamical mean field approximation. 
The two-photon absorption (TPA) and the third-harmonic generation (THG) 
spectra are calculated, and 
the classification by the types of 
coupling to external fields 
shows different behavior from conventional semiconductors. 
The direct transition terms are predominant 
both in the TPA and THG spectra, and 
the importance of taking all types of interaction with 
the external field into account is 
illustrated in connection with the THG spectrum 
and dc Kerr effect. 
The dependences of the TPA and THG spectra 
on the Coulomb interaction indicate a scaling relation. 
We apply this relation to the quantitative evaluation and 
obtain results comparable to those of experiments. 
}

\kword{%
nonlinear optics, two-photon absorption, 
third-harmonic generation, dynamical mean field, 
electron correlation, Mott insulator 
}

\begin{document}

\maketitle

\section{Introduction}

Several nonlinear optical responses have been 
observed in Mott insulators; 
the two-photon absorption (TPA)~\cite{ogasawara,ashida}, 
the third-harmonic generation (THG)~\cite{kishida2,ono} 
and the electroreflectance spectroscopy~\cite{kishida}. 
A notable point is that quasi one-dimensional (1D) Mott insulators 
show large nonlinear responses in these measurements, 
compared with those of conventional semiconductors. 
On the other hand magnitudes of nonlinear responses in 
two-dimensional (2D) systems are comparable to those of 
conventional semiconductors, and then 
the dimensionality dependence of nonlinear 
susceptibilities has also attracted attention 
in Mott insulators.~\cite{ashida,ono}
However this does not mean that 
the 2D system does not need 
an explanation, because the origins of the optical gap 
in the band insulators and Mott insulators 
are different from each other and the theory of 
conventional semiconductors does not apply to Mott insulators. 
There exists detailed comparison between 
experiments and theory 
in conventional semiconductors.~\cite{sheikbahae}  
By contrast, optical nonlinearities in Mott insulators 
have not yet been understood to that level. 

In the previous paper we derived the general formulation 
of the nonlinear optical susceptibility based on 
Green's function, and 
applied this to a calculation of the TPA spectrum 
of antiferromagnetic insulators with the 
Hartree-Fock approximation.~\cite{jujo} 
The dimensionality dependence of nonlinear susceptibilities 
was investigated and a semiquantitative estimation was made there. 
This calculation fails to include the damping effect, 
and the divergence arises at the band edge. 
This makes a quantitative estimation difficult and 
it is done with the averaged spectrum. 
(The damping effect is also important 
due to the experimental fact that 
the response time of Mott insulators is very fast.~\cite{ogasawara})
Other approaches on nonlinear optical responses 
have been made with use of the numerical diagonalization 
method on small-sized systems.~\cite{mizuno} 
This calculation consists of the discrete levels and 
dipole moments between them, and requires an artificial 
damping term. 
Although the qualitative reproduction of 
the dimensionality dependence is made with this method, 
even rough estimation of magnitudes 
of nonlinear susceptibilities is not attempted. 

In this paper we study nonlinear susceptibilities of 
Mott insulators with the dynamical mean field 
approximation on the basis of the general formulation 
developed in ref.~\cite{jujo} 
The damping effect is naturally included within this method. 
We calculate the THG spectrum and dc Kerr effect 
as well as the TPA spectrum. 
It is shown that the direct transition term predominates 
in the TPA and THG spectra. 
This is not the case in the dc Kerr effect, but 
all types of processes are important in the same degree to 
form the oscillating structure. 
The scaling relations of the optical responses are derived, 
and the linear and nonlinear responses are proportional 
to the inverse of the square and the fourth power of the 
energy gap, respectively. 
According to this relation we obtain quantitative results of 
the TPA and THG spectra, which is comparable 
to experiments in the case that the value of 
the Coulomb interaction is somewhat larger than 
that of the bandwidth. 

We present our formulation for calculation in \S 2, 
and the results are shown in \S 3. Several 
vertex corrections to the nonlinear susceptibilities 
are considered in Appendix. 
We set $\hbar={\rm c}=1$ and the electric charge ${\rm e}$ 
is not written explicitly. 
These are restored in quantitative calculations. 

\section{Formulation}

Firstly we show how the Mott insulating state is described 
in our calculation. 
We apply the dynamical mean field approximation (DMFA)
to the single-band Hubbard model, 
\begin{equation}
 {\cal H}=\sum_{<ij>\sigma}t_{ij}
(c^{\dagger}_{i\sigma}c_{j\sigma}+c^{\dagger}_{j\sigma}c_{i\sigma})
+U\sum_i n_{i\uparrow}n_{i\downarrow}. 
\end{equation}
($t_{ij}$ is the transfer integral and $U$ indicates 
the on-site Coulomb interaction.)
We do not use the notation `theory' 
which is usually used in the dynamical mean field theory (DMFT), 
but adopt `approximation' 
because we do not consider the limit of the dimensionality $d\to\infty$. 
This implies the following. 
In DMFT the effective single-site action, 
\begin{equation}
 S_{\rm eff}=-\int_0^{\beta}{\rm d}\tau\int_0^{\beta}{\rm d}\tau'
\sum_{\sigma}c^{\dagger}_{\sigma}(\tau)
\mathcal{G}_0^{-1}(\tau-\tau')c_{\sigma}(\tau')
+U\int_0^{\beta}{\rm d}\tau n_{\uparrow}(\tau)n_{\downarrow}(\tau)
\end{equation}
is derived in the large dimension limit, 
$d\rightarrow\infty$.~\cite{georges}
($\beta=1/T$ and $T$ is the temperature.)
In our case we use this effective action 
in arbitrary lattice systems. 
This means we neglect the higher-order terms 
of the transfer integral other than the first term of $S_{\rm eff}$. 
This is the reason why we use DMFA instead of DMFT. 
In this case 
we do not need to scale the transfer integral 
by the factor of $1/\sqrt{d}$. 

Other processes in the calculation are the same as in DMFT. 
The self-energy is calculated with $S_{\rm eff}$ 
as the functional of $\mathcal{G}_0$, $\Sigma[\mathcal{G}_0]$. 
The Weiss function $\mathcal{G}_0$ is calculated by the 
following relation, 
\begin{equation}
 \mathcal{G}_0^{-1}(\epsilon_{n})
={\rm i}\epsilon_n+\mu_0-G^{(0)}(\epsilon_n)
\end{equation}
and 
\begin{equation}
 G^{(0)}(\epsilon_n)=\sum_k\xi_k^2G_k(\epsilon_n)
-[\sum_k\xi_kG_k(\epsilon_n)]^2/\sum_kG_k(\epsilon_n), 
\end{equation}
with Green's function, 
\begin{equation}
 G_k(\epsilon_n)=\frac{1}{{\rm i}\epsilon_n-\xi_k+\mu-\Sigma(\epsilon_n)}.
\end{equation}
(Here $\epsilon_n=\pi T(2n-1)$ and $n$ is integer.)
These functions are self-consistently determined, and 
the chemical potential $\mu$, $\mu_0$ is fixed 
by the condition, 
$n_{\sigma}[G]=n_{\sigma}[\mathcal{G}_0]=1/2$
(this sets the system to be half-filled). 
We make another approximation to solve $S_{\rm eff}$. 
We calculate the self-energy within 
the second order perturbation,~\cite{zhang} 
\begin{equation}
 \Sigma(\epsilon_n)=-U^2T^2\sum_{n',l}
\mathcal{G}_0(\epsilon_{n'})\mathcal{G}_0(\epsilon_{n'}+\omega_l)
\mathcal{G}_0(\epsilon_{n}-\omega_l). 
\end{equation}
We use the following dispersion relation, 
\begin{equation}
 \xi_k=-2t({\rm cos}k_x+\eta{\rm cos}k_y)+4t'\eta{\rm cos}k_x{\rm cos}k_y.
\end{equation}
In numerical calculations below we put $t=1$ and 
fix the next-nearest-neighbor 
hopping $t'=0.2$ (results do not change if we vary $t'$ moderately). 
We vary $\eta$ as the dimensionality parameter from the 2D 
$\eta=1.0$ to the quasi 1D $\eta=0.1$. 

Next we present the formulation of the nonlinear 
optical response functions. 
The third-order nonlinear susceptibility is determined by, 
\begin{equation}
 \chi^{(3)}(\omega,\omega_1,\omega_2)=
\frac{K^{(3)}(\omega,\omega',\omega'')}{\omega\omega_1\omega_2\omega_3}.
\end{equation}
(The definitions of $\chi^{(3)}$ and $K^{(3)}$ are given 
in ref.~\cite{jujo}) 
Here, $\omega=\omega_1+\omega_2+\omega_3$, $\omega'=\omega_2+\omega_3$ 
and $\omega''=\omega_3$. 
$\omega_{1}$, $\omega_{2}$ and $\omega_{3}$ 
are frequencies of the external fields and 
take different values depending on various methods of measurements. 
$K^{(3)}$ is classified by the types of the coupling to the 
external fields as follows, 
\begin{equation}
 K^{(3)}(\omega,\omega',\omega'')
=K^{(3)}_{<j4>}+K^{(3)}_{<j3>}+K^{(3)}_{<j2>}. \label{eq:k3}
\end{equation}
Each term is written as, 
\begin{equation}
 K^{(3)}_{<j4>}=\frac{2}{3!}
\sum_k\int\frac{{\rm d}\epsilon}{2\pi}v_k^4\sum_{<i,j>}
(G_a^RG_i^RG_j^RT_b+G_a^RG_i^RT_jG_b^A
+G_a^RT_iG_j^AG_b^A+T_aG_i^AG_j^AG_b^A), \label{eq:k3j4}
\end{equation}
\begin{equation}
\begin{split}
 K^{(3)}_{<j3>}=&\frac{2}{3!}\sum_k
\int\frac{{\rm d}\epsilon}{2\pi}\frac{\partial v_k}{\partial k}v_k^2
[\sum_i(G_a^RG_i^RT_b+G_a^RT_iG_b^A+T_aG_i^AG_b^A) 
+\sum_j(G_a^RG_j^RT_b+G_a^RT_jG_b^A+T_aG_j^AG_b^A) \\
&+\sum_{<i,j>}(G_i^RG_j^RT_b+G_i^RT_jG_b^A+T_iG_j^AG_b^A)], \label{eq:k3j3}
\end{split}
\end{equation}
and 
\begin{equation}
\begin{split}
 K^{(3)}_{<j2>}=&\frac{2}{3!}\sum_k
\int\frac{{\rm d}\epsilon}{2\pi}
[\left(\frac{\partial v_k}{\partial k}\right)^2
\sum_i(G_i^RT_b+T_iG_b^A)
+\frac{\partial^2 v_k}{\partial k^2}v_k
\sum_j(G_j^RT_b+T_jG_b^A) \\
&+\frac{\partial^2 v_k}{\partial k^2}v_k
(G_a^RT_b+T_aG_b^A-G_b^RT_b-T_bG_b^A)]. \label{eq:k3j2}
\end{split}
\end{equation}
Here $G^{R,A}_x=G^{R,A}_k(\epsilon_x)$ ($R$ and $A$ mean the 
retarded and advanced, respectively), 
$T_x={\rm tanh}(\epsilon_x/2T){\rm Im}G^{R}_k(\epsilon_x)$ and 
$v_k=\partial \xi_k/\partial k$. 
$\epsilon_x=\epsilon+\omega_x$ and 
$\omega_a=\omega_1+\omega_2+\omega_3$, $\omega_b=0$, 
$\omega_i=\omega_1+\omega_2$, $\omega_1+\omega_3$ or $\omega_2+\omega_3$, 
$\omega_j=\omega_1$, $\omega_2$ or $\omega_3$. 
The diagrammatic representations are given in 
Fig. 1 of ref.~\cite{jujo}
; Fig. 1(a), (b) and (c,d,e) for $K^{(3)}_{<j4>}$, 
$K^{(3)}_{<j3>}$ and $K^{(3)}_{<j2>}$, respectively. 
In this formulation vertex corrections are omitted, and 
these are discussed in Appendix. 

\section{Results}

\subsection{The analysis of spectrum}

The numerical results shown below are 
calculated with eqs. (\ref{eq:k3j4},\ref{eq:k3j3},\ref{eq:k3j2}). 
The vertex corrections are not included, 
which are small compared to these terms as indicated in Appendix. 

\begin{figure}
\includegraphics[width=8.5cm]{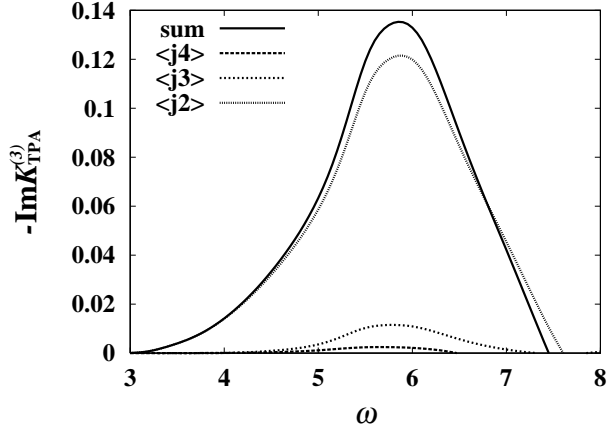}
\caption{The decomposition of ${\rm Im}K^{(3)}$ 
in the case of the TPA spectrum. $U=12$ and $\eta=0.2$ 
'sum' indicates the sum of three terms.} 
\label{fig:1}
\end{figure}

The decomposition of ${\rm Im}K^{(3)}$ to 
${\rm Im}K^{(3)}_{<j4>}$, ${\rm Im}K^{(3)}_{<j3>}$ 
and ${\rm Im}K^{(3)}_{<j2>}$ in the case of 
the TPA spectrum ($\omega_1=-\omega_2=\omega_3=\omega$)
is shown in Fig.~\ref{fig:1}. 
(We fix the temperature $T=0.036$ hereafter, and 
this parameter is not considered to be important 
because of $\omega,U,t\gg T$.) 
The predominance of $K^{(3)}_{<j2>}$ over $K^{(3)}_{<j4>}$ 
and $K^{(3)}_{<j3>}$ is peculiar to Mott insulators, 
in contrast with conventional semiconductors 
where $K^{(3)}_{<j2>}$ vanishes except 
for the self-transition.~\cite{yee,wherrett} 
The existence of $K^{(3)}_{<j2>}$ in the TPA spectrum 
depends on the origin of the gap,~\cite{jujo} 
and the difference in magnitude of three ${\rm Im}K^{(3)}$ 
is understood by writing expressions explicitly as follows. 
\begin{equation}
\begin{split}
 {\rm Im}K^{(3)}_{<j4>}(\omega,0,\omega)
&=\frac{4}{3}\sum_k\int\frac{{\rm d}\epsilon}{2\pi}v_k^4
[
\left(
{\rm tanh}\frac{\epsilon-\omega}{2T}-{\rm tanh}\frac{\epsilon+\omega}{2T}
\right)
I_{\epsilon+\omega}R_{\epsilon}I_{\epsilon-\omega}R_{\epsilon} \\
&+
\left(
{\rm tanh}\frac{\epsilon+\omega}{2T}-{\rm tanh}\frac{\epsilon}{2T}
\right)
I_{\epsilon+\omega}I_{\epsilon}
(I_{\epsilon+\omega}I_{\epsilon}-R_{\epsilon+\omega}R_{\epsilon}
-R_{\epsilon+2\omega}R_{\epsilon+\omega}-R_{\epsilon}R_{\epsilon-\omega})
]. 
\end{split}
\end{equation}
\begin{equation}
\begin{split}
 {\rm Im}K^{(3)}_{<j3>}(\omega,0,\omega)
&=\frac{4}{3}\sum_k\int\frac{{\rm d}\epsilon}{2\pi}
\frac{\partial v_k}{\partial k}v_k^2
[
\left(
{\rm tanh}\frac{\epsilon-\omega}{2T}-{\rm tanh}\frac{\epsilon+\omega}{2T}
\right)
I_{\epsilon+\omega}R_{\epsilon}I_{\epsilon-\omega} \\
&+
\left(
{\rm tanh}\frac{\epsilon}{2T}-{\rm tanh}\frac{\epsilon+\omega}{2T}
\right)
I_{\epsilon+\omega}I_{\epsilon}
(R_{\epsilon+\omega}+R_{\epsilon}
+R_{\epsilon+2\omega}/2+R_{\epsilon-\omega}/2)
]. 
\end{split}
\end{equation}
\begin{equation}
\begin{split}
 {\rm Im}K^{(3)}_{<j2>}(\omega,0,\omega)
=\frac{1}{3}\sum_k\int\frac{{\rm d}\epsilon}{2\pi}
&
[\left(\frac{\partial v_k}{\partial k}\right)^2
\left(
{\rm tanh}\frac{\epsilon-\omega}{2T}-{\rm tanh}\frac{\epsilon+\omega}{2T}
\right)
I_{\epsilon+\omega}I_{\epsilon-\omega} \\
+&
2\frac{\partial^2 v_k}{\partial k^2}v_k
\left(
{\rm tanh}\frac{\epsilon}{2T}-{\rm tanh}\frac{\epsilon+\omega}{2T}
\right)
I_{\epsilon+\omega}I_{\epsilon}
]. 
\end{split}
\end{equation}
Here $I_{\epsilon}={\rm Im}G^R_k(\epsilon)$ 
and $R_{\epsilon}={\rm Re}G^R_k(\epsilon)$. 
We consider the case of $\omega\simeq U/2$ in the TPA spectrum. 
These expressions indicate that 
the second terms of these three equations are small 
due to the factor 
$I_{\epsilon+\omega}I_{\epsilon}
={\rm Im}G^R_k(\epsilon+\omega){\rm Im}G^R_k(\epsilon)$. 
(If one of ${\rm Im}G^R$ takes large values, the other has 
small values owing to the absence of the spectrum.) 
Then we consider the first terms in these expressions. 
The existence of $R_{\epsilon}$ is the reason for 
the smallness of $K^{(3)}_{<j4>}$ and $K^{(3)}_{<j3>}$ 
compared with the direct transition term $K^{(3)}_{<j2>}$. 
The former two cases includes virtually excited states 
in the optical process, and $R_{\epsilon}$ expresses 
this excitation. 
Although $I_{\epsilon+\omega}I_{\epsilon-\omega}
={\rm Im}G^R_k(\epsilon+\omega){\rm Im}G^R_k(\epsilon-\omega)$ 
can take large values around $\epsilon\simeq 0$, 
$R_{\epsilon}$ is roughly proportional to $1/U$ in this region 
and is small. This explains results of Fig.~\ref{fig:1}. 

\begin{figure}
\includegraphics[width=8.5cm]{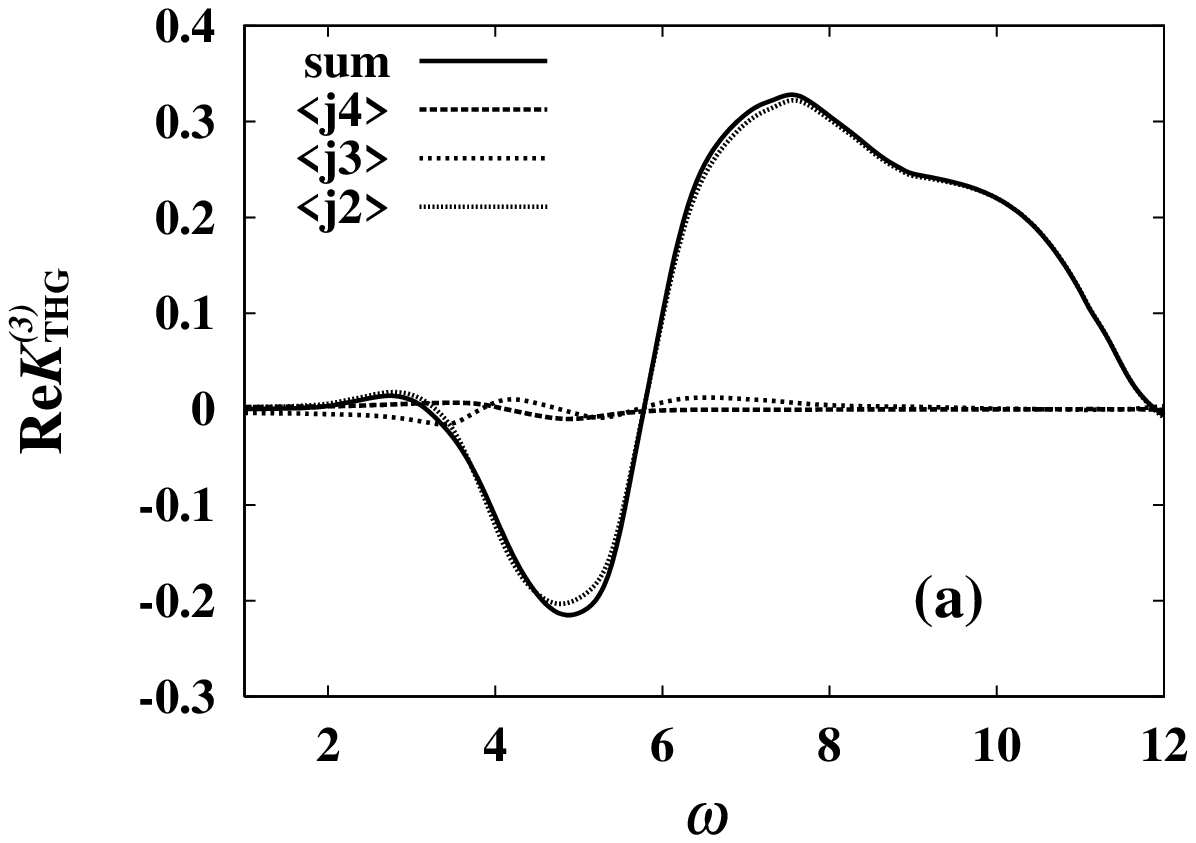}
\includegraphics[width=8.5cm]{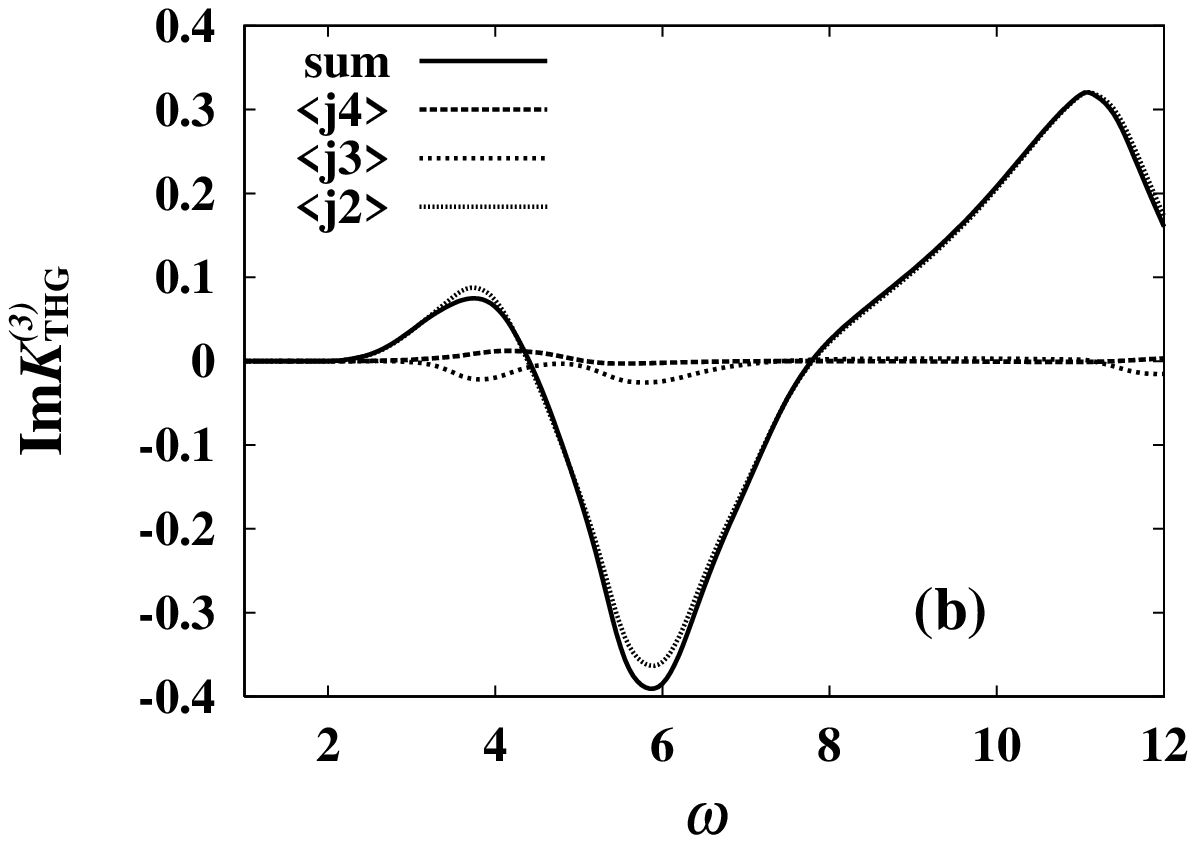}
\caption{The decomposition of (a) the real and (b) imaginary part 
of $K^{(3)}$ in the case of the THG spectrum. $U=12$ and $\eta=0.2$. 
'sum' indicates the sum of three terms.}
\label{fig:2}
\end{figure}

\begin{figure}
\includegraphics[width=8.5cm]{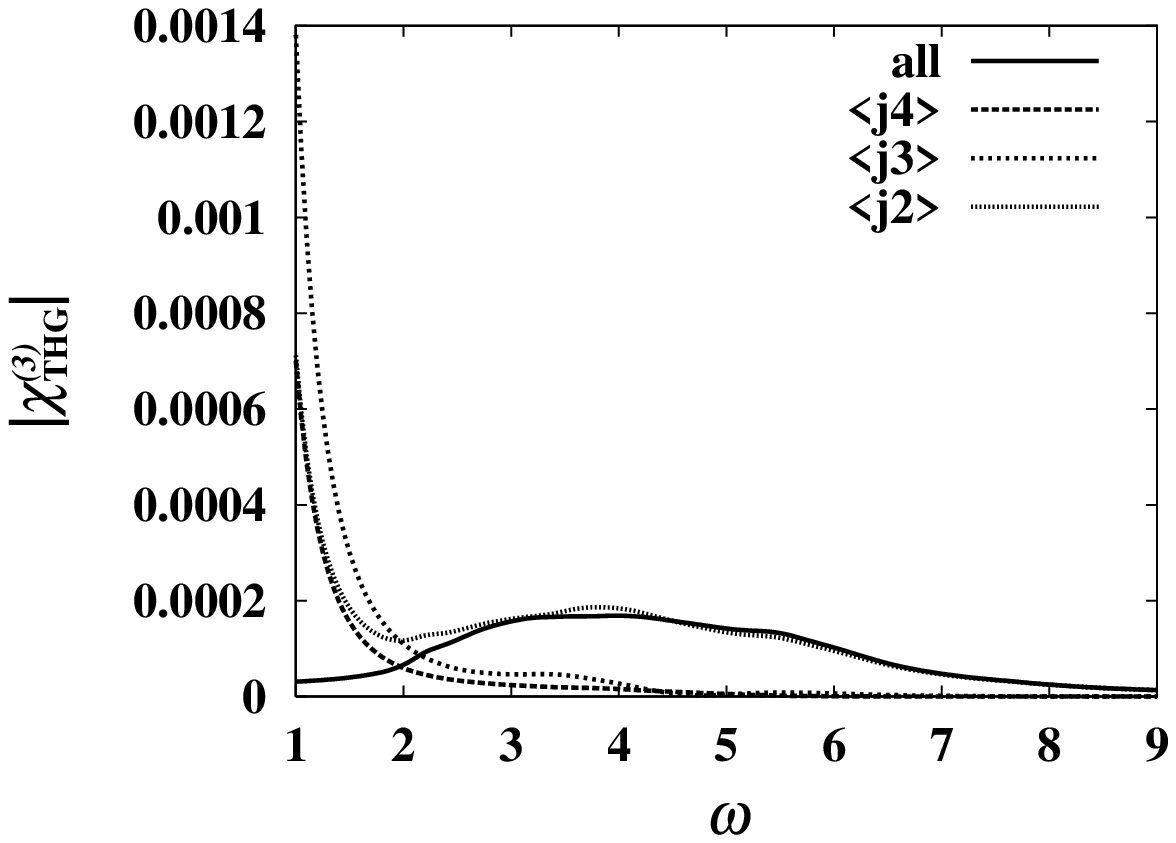}
\caption{The decomposition of 
$|\chi^{(3)}_{\rm THG}|$ in the case of the THG spectrum. 
$|\chi^{(3)}_{<j4>}|$, $|\chi^{(3)}_{<j3>}|$ 
and $|\chi^{(3)}_{<j2>}|$ are calculated with 
$K^{(3)}_{<j4>}$, $K^{(3)}_{<j3>}$ 
and $K^{(3)}_{<j2>}$, respectively. 
'all' means 
 $|\chi^{(3)}_{<j4>}+\chi^{(3)}_{<j3>}+\chi^{(3)}_{<j2>}|$. 
$U=12$ and $\eta=0.2$. 
}
\label{fig:3}
\end{figure}

The decomposition of $K^{(3)}$ to 
$K^{(3)}_{<j4>}$, $K^{(3)}_{<j3>}$ 
and $K^{(3)}_{<j2>}$ in the case of 
the THG spectrum ($\omega_1=\omega_2=\omega_3=\omega$)
is shown in Fig.~\ref{fig:2}. 
The predominance of $K^{(3)}_{<j2>}$ over 
$K^{(3)}_{<j4>}$ and $K^{(3)}_{<j3>}$ 
is the same as the case of the TPA spectrum, 
and the reason for this is also the same. 
(Here we consider the case of $\omega\simeq U/3$.) 
If we write the expressions of $K^{(3)}$ explicitly, 
we can find that the factor like 
${\rm Im}G^R_k(\epsilon+3\omega){\rm Im}G^R_k(\epsilon)$ 
exists in $K^{(3)}_{<j2>}$. 
Then $K^{(3)}_{<j2>}$ takes larger values than 
the other two terms, which include 
the nonresonant $R_{\epsilon}$ term. 
In the THG spectrum 
the existence of the real part ${\rm Re}\chi^{(3)}$ 
makes it inevitable to calculate 
all three terms of $K^{(3)}$ consistently, 
especially for small $\omega$. 
If we calculate $|\chi^{(3)}_{\rm THG}|$ only with 
$K^{(3)}_{<j4>}$, $K^{(3)}_{<j3>}$ or $K^{(3)}_{<j2>}$ separately, 
each of $|\chi^{(3)}_{<j4,j3,j2>}|$ diverges at 
small $\omega$ as shown in Fig.~\ref{fig:3}. 
The cancellation among three $K^{(3)}_{<j4,j3,j2>}$ 
occurs at small $\omega$, and we obtain convergence 
only if the summation of these three terms is taken. 
(This cancellation is the nonlinear analogue of 
that between the paramagnetic and diamagnetic terms 
in the linear response. 
It is unaffected by vertex corrections 
owing to the absence of the momentum-dependence 
in the self-energy.) 
This shows the importance of taking all three terms 
into account. 
The convergent behavior is related to that of 
the Drude weight, which is defined as 
$D:=\pi\omega{\rm Im}\sigma|_{\omega\to 0}$ 
($\sigma$ is the conductivity) 
and $D=0$ for $T\to 0$ in insulators.~\cite{kohn} 
The nonlinear correction is written as 
${\rm Im}\sigma^{(3)}=-{\rm Re}K^{(3)}/\omega^3$. 
Therefore the nonlinear correction to the Drude weight would be 
divergent if ${\rm Re}K^{(3)}$ took finite values. 
(Strictly speaking, the term which is proportional 
to ${\rm exp}(-E_g/T)$ ($E_g$ is the energy gap) 
remains as in the linear response, 
but this is vanishingly small for $E_g\gg T$.) 

\begin{figure}
\includegraphics[width=8.5cm]{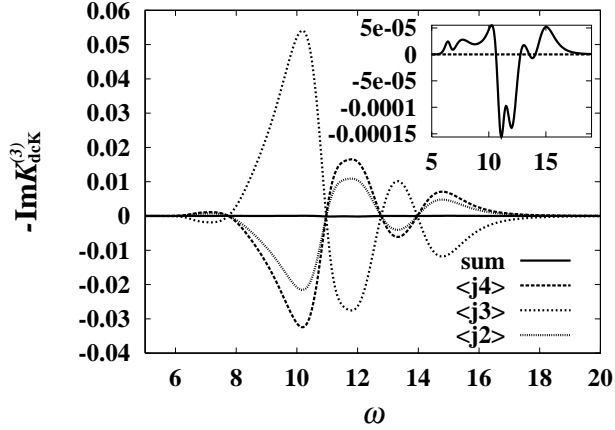}
\caption{The decomposition of 
${\rm Im}K^{(3)}(\omega,0,-\Delta\omega)$. 
$\Delta\omega=0.05$, $U=12$ and $\eta=0.2$. 
'sum' indicates the sum of three terms and 
the inset shows this result separately because of 
the difference in scales.}
\label{fig:4}
\end{figure}

The decomposition of ${\rm Im}K^{(3)}$ to 
${\rm Im}K^{(3)}_{<j4>}$, ${\rm Im}K^{(3)}_{<j3>}$ 
and ${\rm Im}K^{(3)}_{<j2>}$ in an approximate case of 
the dc Kerr effect ($\omega_1=\omega$, $\omega_2=-\omega_3=\Delta\omega$)
is shown in Fig.~\ref{fig:4}. 
(It should be $\Delta\omega\to 0$, but we apply the finite 
difference to $\chi^{(3)}=-K^{(3)}/(\omega^2\Delta\omega^2)$.) 
In contrast to the above two cases, 
all of $K^{(3)}_{<j4>}$, $K^{(3)}_{<j3>}$ and $K^{(3)}_{<j2>}$ 
contributes to $K^{(3)}$ in the same degree. 
The reason for this is that ${\rm Re}G^R_k(\epsilon)$ 
does not necessarily locate at the nonresonant state, 
which is understood by writing the set of frequencies; 
$(\omega_i,\omega_j)=(0,\Delta\omega)$, $(0,-\Delta\omega)$, 
$(\omega+\Delta\omega,\Delta\omega)$, $(\omega+\Delta\omega,\omega)$, 
$(\omega-\Delta\omega,-\Delta\omega)$, $(\omega-\Delta\omega,\omega)$. 
As shown in the inset 
the summation of these three terms is 
smaller than each of them by two orders of magnitude. 
All these terms are required to reproduce 
the characteristic oscillating structure 
similar to that observed 
in the electroreflectance spectroscopy. 

It is known that sum rules 
hold in the nonlinear response.~\cite{peiponen,bassani} 
The relation, 
$\int_0^{\infty}\omega
\epsilon_2^{\rm NL}(\omega,-\omega',\omega'){\rm d}\omega=0$ 
holds for the TPA spectrum. 
Here, $\epsilon_2^{\rm NL}(\omega,-\omega',\omega')$
is the imaginary part 
of the complex dielectric function. 
If we treat the above three terms of $K^{(3)}$ separately, 
we will violate this relation. 
The appearance of the oscillating structure in 
the dc Kerr effect as shown above is another example 
of the necessity to consider all these terms in $K^{(3)}$ 
($\omega'=\Delta\omega$ in this case). 
A previous calculation do not take 
these terms into account properly.~\cite{jafari}
They neglect the predominant term $K^{(3)}_{<j2>}$, 
and also fail to treat the divergence at small frequency region 
carefully in a calculation of the real part of $\chi^{(3)}_{\rm THG}$ 

\subsection{The dependences of 
nonlinear susceptibilities on the Coulomb interaction}

\begin{figure}
\includegraphics[width=8.5cm]{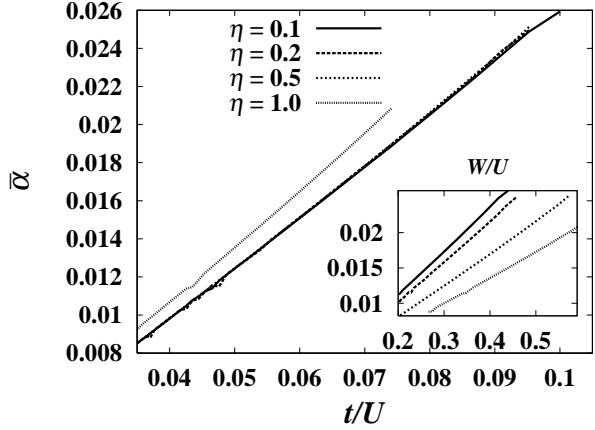}
\caption{The dependence of 
$\bar{\alpha}=
\int\omega{\rm Im}\chi^{(1)}(\omega){\rm d}\omega$ on $t/U$. 
The inset shows the dependence of the same quantities on $W/U$.}
\label{fig:5}
\end{figure}

We show the dependences of the nonlinear 
susceptibilities on $U/t$, $U/W$ and $\eta$. 
(Here $W$ is the bare bandwidth 
and is a function of $t'$ and $\eta$.) 
The dependence of 
the integral of the linear absorption spectrum 
($\bar{\alpha}=\int\omega{\rm Im}\chi^{(1)}(\omega){\rm d}\omega$) 
on $t/U$, $W/U$ with several values of $\eta$ 
is shown in Fig.~\ref{fig:5}.  
(The value of $U$ at which the Mott transition occurs 
depends on $\eta$, and these are 
$U\simeq 9.1,9.2,10.2,13.1$ for $\eta=0.1,0.2,0.5,1.0$, respectively.)
The relation $\bar{\alpha}\propto 1/U$ 
holds, which is consistent with the sum rule for 
the linear absorption.~\cite{baeriswyl} 
If we put the lattice constant $a=5$ \AA  and 
$U=2$ eV (The reason why we take this value is that 
the linear absorption spectrum in experiments 
peaks around this energy and in our simple model 
the spectrum always has the peak around $U$.), we get 
$\alpha^{(1)}|_{\rm peak}\simeq 0.98,0.76 \times 10^{5}$ 
${\rm cm}^{-1}$ at $U=13.5$ 
for $\eta=0.1,1.0$, respectively 
(here $\alpha^{(1)}(\omega)=4\pi\omega{\rm Im}\chi^{(1)}(\omega)/{\rm c}$ 
and ${\rm c}$ is the velocity of light which is written explicitly 
for the quantitative estimation). 
These are almost comparable to the results of experiments 
which are $\alpha^{(1)}|_{\rm peak}\simeq 4,1 \times 10^{5}$ 
${\rm cm}^{-1}$ in quasi 1D and 2D systems, respectively.~\cite{ashida} 
The relation $\omega{\rm Im}\chi^{(1)}\propto 1/U$ indicates 
that we expect a moderate enhancement 
of $\alpha^{(1)}$ for smaller $U$. 

\begin{figure}
\includegraphics[width=8.5cm]{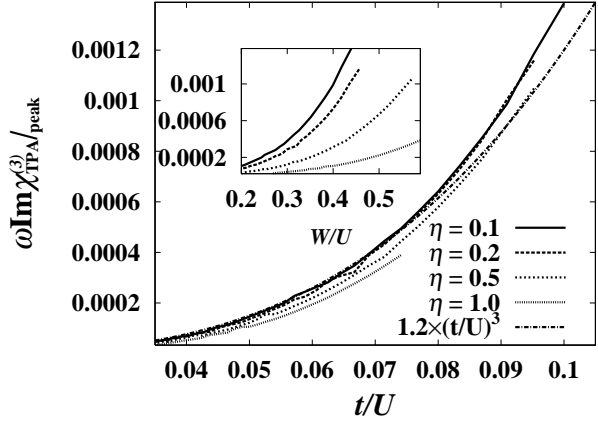}
\caption{The dependence of 
$\omega{\rm Im}\chi^{(3)}_{\rm TPA}|_{\rm peak}$ on $t/U$. 
The inset shows the dependence of the same quantities on $W/U$.}
\label{fig:6}
\end{figure}

The dependences of the peak of the TPA spectrum 
(${\rm Im}\chi^{(3)}_{\rm TPA}$ multiplied by $\omega$) 
on $t/U$ and $W/U$ are shown in Fig.~\ref{fig:6}. 
The relation 
$\omega{\rm Im}\chi^{(3)}_{\rm TPA} \propto 1/U^3$ 
holds approximately. 
(It deviates slightly from $1/U^3$ for smaller U, 
and the results are rather proportional to $1/U^{3.5}$. 
This is because the peaks of the TPA spectrum shift 
to lower energies.) 
If we put the lattice constant $a=5$ \AA and 
$U=2$ eV, we get 
${\rm Im}\chi^{(3)}_{\rm TPA}\simeq 0.0155,0.0133\times 10^{-9}$ esu 
at $U=13.5$ for $\eta=0.1,1.0$, respectively. 
If we extrapolate the relation 
${\rm Im}\chi^{(3)}_{\rm TPA} \propto 1/U^4$ 
for smaller $U$, we will obtain  
${\rm Im}\chi^{(3)}_{\rm TPA}\simeq 1.0, 0.1\times 10^{-9}$ esu 
at $U=4.76$, $\eta=0.1$ and $U=8.15$, $\eta=1.0$, respectively. 

\begin{figure}
\includegraphics[width=8.5cm]{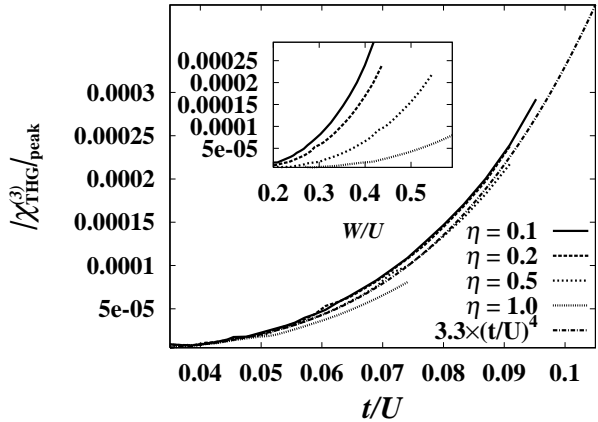}
\caption{The dependences of $|\chi^{(3)}_{\rm THG}|_{\rm peak}$ on $t/U$. 
The inset shows the dependence of the same quantities on $W/U$.}
\label{fig:7}
\end{figure}

The dependences of the peak of the THG spectrum on $t/U$ and $W/U$ 
are shown in Fig.~\ref{fig:7}. 
The relation $|\chi^{(3)}_{\rm THG}|\propto 1/U^4$ holds. 
If we set parameters same as above to evaluate 
$|\chi^{(3)}_{\rm THG}|$ quantitatively, we get 
$|\chi^{(3)}_{\rm THG}|\simeq 0.0217, 0.0162\times10^{-9}$ esu 
at $U=13.5$ for $\eta=0.1,1.0$, respectively. 
If we assume that 
the relation $|\chi^{(3)}_{\rm THG}|\propto 1/U^4$ holds 
for smaller $U$, we will obtain 
$|\chi^{(3)}_{\rm THG}|\simeq 1.0,0.1\times10^{-9}$ esu, 
at $U=5.18$, $\eta=0.1$ and $U=8.56$, $\eta=1.0$, respectively. 

These results indicate that 
the dependence of the susceptibility on $\eta$ 
is rather weak, at least with $t/U$ fixed. 
On the other hand it is strongly dependent on $\eta$ 
in the case of $W/U$ fixed, and this is because 
the bandwidth $W$ is a function of $\eta$. 
The experimental results indicate that 
$\chi^{(3)}\simeq 1.0,0.1\times10^{-9}$ esu 
for quasi 1D and 2D systems, respectively.~\cite{ashida,ono} 
Our calculation shows that 
it is possible to obtain $\chi^{(3)}$ 
comparable to those of experiments in the case 
of $U\gtrsim W$ (actually $W=4.4$ and $8.0$ for 
$\eta=0.1$ and $1.0$, respectively). 
However this is based on the condition that 
we can extrapolate scaling relations for smaller $U$, 
and we discuss this point in \S 4. 
We find that the dependences of $\chi$ on $t'$ is weak 
with the moderate variation of $t'$. 

\begin{figure}
\includegraphics[width=8.5cm]{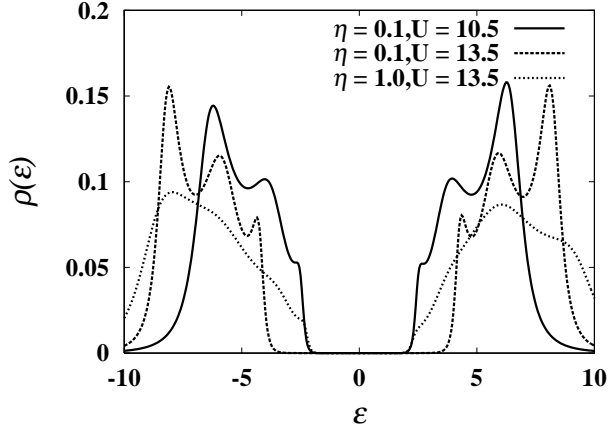}
\caption{The density of states 
$\rho(\epsilon)=-\sum_k{\rm Im}G^R_k(\epsilon)/\pi$ 
with several values of $U$ and $\eta$.}
\label{fig:8}
\end{figure}

\begin{figure}
\includegraphics[width=8.5cm]{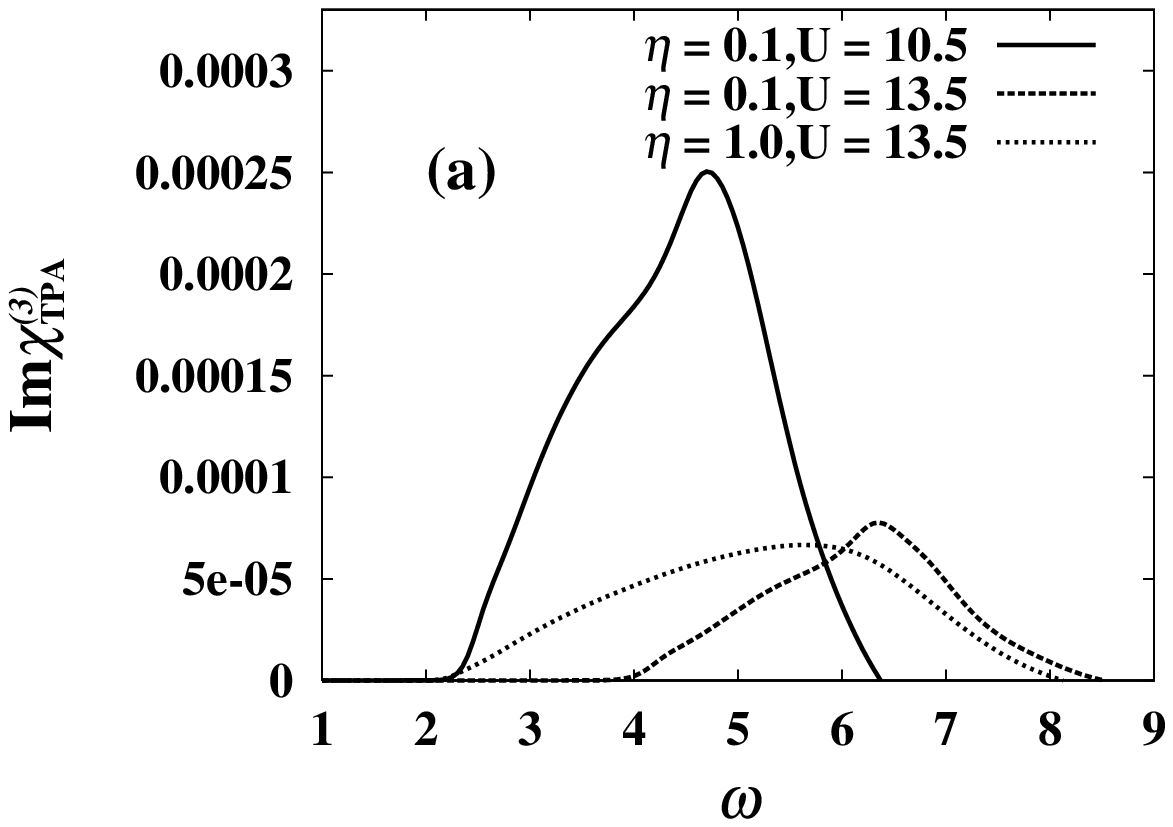}
\includegraphics[width=8.5cm]{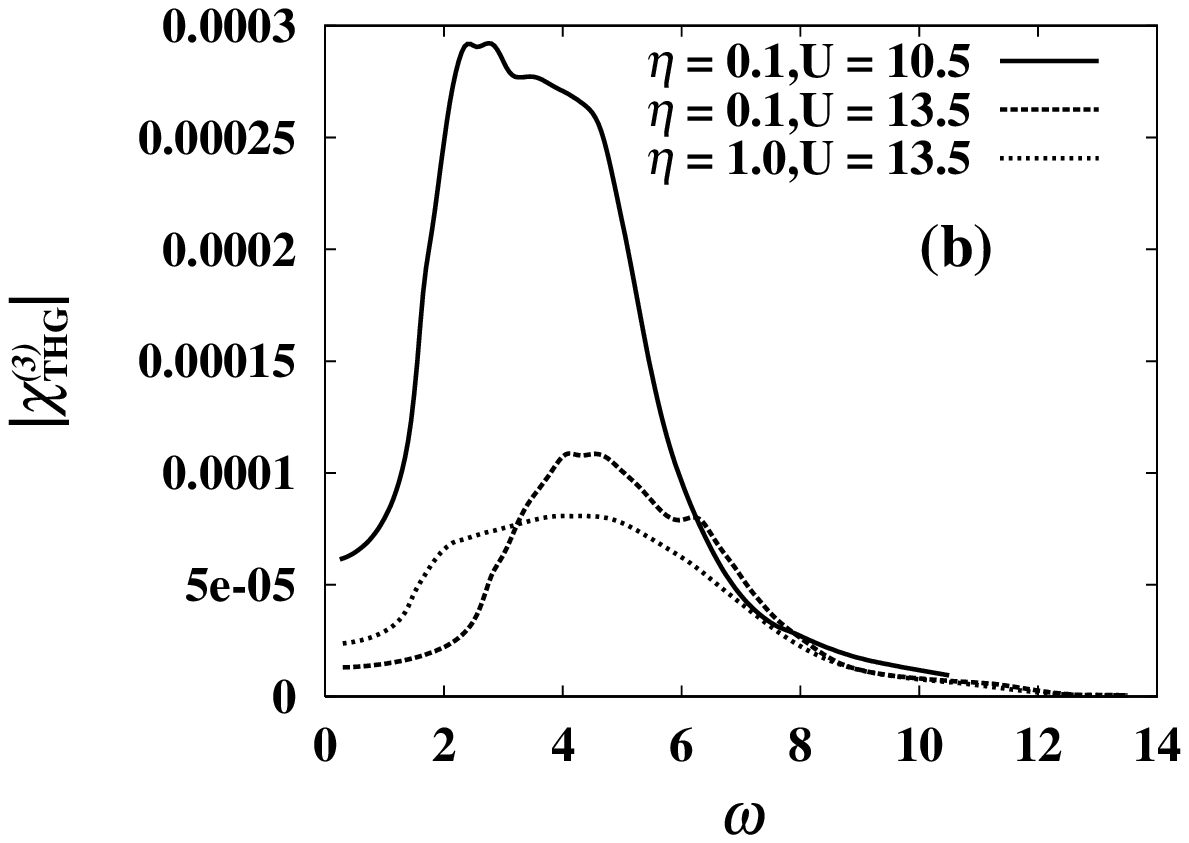}
\caption{(a) ${\rm Im}\chi^{(3)}_{\rm TPA}$ and (b) $|\chi^{(3)}_{THG}|$ 
with several values of $U$ and $\eta$.}
\label{fig:9}
\end{figure}

In experiments 
the nonlinear susceptibility in the quasi 1D system 
is one order of magnitude larger than that in the 2D system. 
Our result does not show so much difference between 
$\eta=0.1$ and $\eta=1.0$ with fixed $t/U$. 
Although the improvement should be done on DMFA 
especially in quasi 1D systems, 
this is partly explained by the behavior of the density 
of states, which is shown in Fig.~\ref{fig:8}.  
The experiment on the linear absorption spectrum 
indicates that the band-edges of the spectrum are 
almost same in both systems. 
This means that the nonlinear susceptibilities 
to be compared should have 
the same band-edge in the density of states. 
Therefore we compare the nonlinear susceptibilities
at $U=10.5,\eta=0.1$ and $U=13.5,\eta=1.0$ as 
an example having such properties. 
The TPA and THG spectra 
are shown in Fig.~\ref{fig:9}. 
According the scaling relation 
$\chi^{(3)}_{\rm TPA,\rm THG}\propto 1/U^4$, 
a slight change of $U$ brings about large variations in 
the nonlinear optical susceptibilities. 
On the other hand the linear absorption spectrum 
does not change considerably because of 
$\chi^{(1)}\propto 1/U^2$. 
Consequently the ratio of $\chi^{(3)}|_{\eta=0.1}$ 
to $\chi^{(3)}|_{\eta=1.0}$ becomes much larger than 
that of $\chi^{(1)}|_{\eta=0.1}$ 
to $\chi^{(1)}|_{\eta=1.0}$, which 
resembles the observations in experiments. 

The scaling relation in semiconductors shows that 
$\chi^{(3)}_{\rm TPA}\propto 1/E_g^4$.~\cite{wherrett,sheikbahae}
($E_g$ is the energy gap.) 
Although this is similar to our result, 
this does not mean that both Mott insulators and 
conventional semiconductors obey the same scaling relation 
because the dominant terms in $\chi^{(3)}$ 
are different between these materials as mentioned in \S 3.1. 
In spite of this fact, 
the difference in the magnitude of 
the nonlinear susceptibility between these materials 
is partly explained as follows. 
For the low dimensional systems 
the gap edge of the density of states is steeper 
than that of more high dimensional systems as 
shown in Fig.~\ref{fig:8}.  
This enhances the magnitude of the optical susceptibility 
in quasi 1D systems, compared to 
that conventional semiconductors. 
 
\section{Summary and Discussion}

We calculate nonlinear optical susceptibilities 
with DMFA on the basis of the general formulation 
of nonlinear response developed in a previous paper. 
The direct transition term is predominant in 
the TPA and THG spectra, which is contrary to 
conventional semiconductors. 
This is because the transition to 
the nonresonant intermediate states gives 
small contribution to $\chi^{(3)}$ due to the strong correlation. 
On the other hand the origin of the band gap in semiconductors 
makes the direct transition negligible in $\chi^{(3)}$. 
In spite of these facts our result shows that 
as a function of the energy gap 
the scaling relation in Mott insulators behaves 
similarly as that of conventional semiconductors. 
A semiquantitative evaluation of nonlinear susceptibilities 
is carried out and shows that results are comparable to 
those of experiments on the condition that the value of 
the Coulomb interaction is somewhat larger than 
the bandwidth. 
The magnitude of ${\rm Im}\chi^{(3)}_{\rm TPA}$ 
and $|\chi^{(3)}_{\rm THG}|$ takes similar values 
with each other, which is also indicated 
by experiments. 
These are not clarified in previous works 
for small systems which are diagonalized numerically. 
The scaling relation based on DMFA also shows that 
the smaller $U$ is favorable to the larger $\chi^{(3)}$ as in 
the Hartree-Fock calculation, which is 
contrary to the scenario of a large optical nonlinearity 
based on the spin-charge separation.~\cite{mizuno} 
(The spin-charge separation holds approximately and is preferred 
at large $U/t$. 
The validity of the spin-charge separation 
as an explanation for the large optical nonlinearity 
can be judged partly from the dependence of 
$\chi^{(3)}$ on parameters like $U/t$.) 

One of our conclusions is dependent on 
the assumption that the scaling relation holds for smaller $U$. 
Here we discuss on this point and a possible modification. 
The main reason why the relations 
$\chi^{(1)}\propto 1/U^2$ and 
$\chi^{(3)}_{\rm TPA,\rm THG}\propto 1/U^4$ hold is as follows. 
By definition $\chi^{(1)}\propto 1/\omega^2$ and 
$\chi^{(3)}_{\rm TPA,\rm THG}\propto 1/\omega^4$. 
This leads to the above $U$-dependences on the condition that 
the $U$-dependences of 
$K^{(1)}$ and $K^{(3)}_{\rm TPA,\rm THG}$ are weak and 
the optical gap scales with $U$. 
The calculation here indicates that this property holds
at least within our approximation. 
However there is some room for improvement 
with respect to the description of the Mott insulator. 
The Hubbard model is considered to have 
the Mott transition at smaller values of $U$ 
than those of a calculation presented here. 
This is the case especially in the model with $\eta=0.0$, 
which is the 1D system and should be 
the Mott insulator even as $U\to 0$.~\cite{lieb} 
The improvement should be done on this point to 
examine the dimensionality dependences and 
the scaling relation for smaller $U$ 
(for example, an expansion to include $k$-dependence 
of the self-energy~\cite{maier}). 

\section*{Acknowledgement}
Numerical computation in this work was carried out at 
the Yukawa Institute Computer Facility. 

\appendix
\section{Vertex corrections}

The correction to vertices $v_k$ and 
$\partial^2 v_k/\partial k^2$ vanishes as in ref.~\cite{khurana,zlatic} 
owing to the inversion symmetry. 
On the other hand it is not known to what extent 
the correction to vertices 
$\partial v_k/\partial k$ and $v_k^2$ 
contributes to $\chi^{(3)}$. 
The diagrams and equations of this type of vertices 
are similar to those of Fig. 1 (f,g,h) and 
\S 3.2 in ref.~\cite{jujo}. 
The vertex correction to the predominant term in $K^{(3)}$ 
is written as, 
\begin{equation}
K^{(3)}_{\rm vc}(\omega_l)=-T^2\sum_{n,n'}
\sum_{k}G_k(\epsilon_n+\omega_l)
\frac{\partial v_k}{\partial k}G_k(\epsilon_n)
\Gamma_(\epsilon_n,\epsilon_n';\omega_l)
\sum_{k'}G_{k'}(\epsilon_{n'}+\omega_l)
\frac{\partial v_{k'}}{\partial k'}G_{k'}(\epsilon_{n'}). 
\end{equation}
Here $\Gamma(\epsilon_n,\epsilon_{n'};\omega_l)$ is the reducible 
four-point vertex. 
If we consider the second-order perturbation term 
as an irreducible four-point vertex $I(\epsilon_n,\epsilon_{n'};\omega_l)$, 
it is written as 
$I(\epsilon_n,\epsilon_{n'};\omega_l)
=2\chi(\epsilon_n-\epsilon_{n'})+\phi(\epsilon_n-\epsilon_{n'})$ 
($\chi(\omega_l)=-U^2T\sum_{k,n}G_k(\epsilon_n+\omega_l)G_k(\epsilon_n)$
and 
$\phi(\omega_l)=-U^2T\sum_{k,n}G_k(\omega_l-\epsilon_n)G_k(\epsilon_n)$). 
From the expression we anticipate that 
the vertex correction is small in the case that 
the dependence of $I(\omega_l)$ on frequency is weak. 
It is because the particle-hole symmetry 
holds approximately. 
This can be verified by the numerical calculation which shows that 
the vertex correction is smaller than ordinary terms 
by two orders of magnitude. 

\begin{figure}
\includegraphics[width=8.5cm]{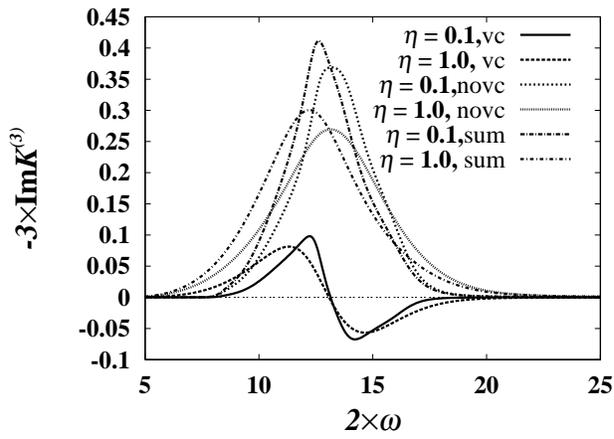}
\caption{The vertex correction to $K^{(3)}_{\rm TPA}$ by the 
nearest-neighbor Coulomb interaction $V$ with the Fock approximation. 
$U=13.5$ and $V=3.0$. 'vc' and 'no vc' means the vertex correction term 
and ${\rm Im}K^{(3)}_{\rm TPA}$ without the vertex correction, and 'sum' means 
the summation of both terms. The vertical and horizontal axes are 
scaled with three times and twice values, respectively.} 
\label{fig:10}
\end{figure}

In contrast to this, 
the nearest-neighbor interaction ($V$) is considered to be 
important in optical responses 
because the excitons can be formed by the 
final-states interaction. 
Therefore we consider the vertex correction by the 
nearest-neighbor interaction. 
The formulation is similar to that of \S 3.2 in ref.~\cite{jujo} 
and we consider only the Fock term with this interaction. 
The vertex correction to the predominant term ${\rm Im}K^{(3)}_{<j2>}$, 
${\rm Im}K^{(3)}_{<j2>}$ itself and the summation of both terms 
of the TPA spectrum are shown in Fig.~\ref{fig:10}. 
The vertex correction shifts the spectrum to lower energy. 
This effect is rather small compared to that of 
antiferromagnetic insulators with the Hartree-Fock approximation 
because the damping effect is included in DMFA. 
Although the value of $V/U$ is not known 
($V/U\simeq 0.22$ in Fig.~\ref{fig:10} 
is considered to be a large value), 
this type of the vertex correction will increase 
the values of $\chi^{(3)}$ of \S 3.2 in some degree.

\end{document}